\documentclass[
 aip,
 amsmath,amssymb,
 reprint,
]{revtex4-2}

\usepackage{graphicx}
\usepackage{dcolumn}
\usepackage{bm}
\usepackage[utf8]{inputenc}
\usepackage[T1]{fontenc}
\usepackage{mathptmx}
\usepackage{etoolbox}
\usepackage{color} 
\usepackage[colorlinks=true, pdfborder={0 0 0}, citecolor=blue, urlcolor=blue]{hyperref}
\usepackage{xcolor}
\usepackage{gensymb}

\usepackage[separate-uncertainty=true,]{siunitx}

\usepackage{booktabs}
 
\usepackage[normalem]{ulem}

\newcommand{\mfzero}{m_\mathrm{F} = 0}
\newcommand{\mftwo}{m_\mathrm{F} = 2}

\newcommand{\zerohk}{0\hbar\mathrm{k}}
\newcommand{\twohk}{2\hbar\mathrm{k}}

\makeatletter
\def\@email#1#2{
 \endgroup
 \patchcmd{\titleblock@produce}
  {\frontmatter@RRAPformat}
  {\frontmatter@RRAPformat{\produce@RRAP{*#1\href{mailto:#2}{#2}}}\frontmatter@RRAPformat}
  {}{}
}
\makeatother

\begin{document}

\preprint{AIP/123-QED}

\title[]{Space magnetometry with a differential atom interferometer}

\author{Matthias Meister\textsuperscript{\dag}}
\affiliation{German Aerospace Center (DLR), Institute of Quantum Technologies, Ulm, Germany.}
\email{Matthias.Meister@dlr.de}

\author{Gabriel Müller\textsuperscript{\dag}}
\affiliation{Leibniz University Hannover, Institute of Quantum Optics, QUEST-Leibniz Research School, Hanover, Germany.}

\author{Patrick Boegel}
\affiliation{Institut für Quantenphysik and Center for Integrated Quantum Science and Technology (IQST), Ulm University, Ulm, Germany.}

\author{Albert Roura}
\affiliation{German Aerospace Center (DLR), Institute of Quantum Technologies, Ulm, Germany.}

\author{Annie Pichery}
\affiliation{Leibniz University Hannover, Institute of Quantum Optics, QUEST-Leibniz Research School, Hanover, Germany.}

\author{David B. Reinhardt}
\affiliation{German Aerospace Center (DLR), Institute of Quantum Technologies, Ulm, Germany.}

\author{Timothé Estrampes}
\affiliation{Leibniz University Hannover, Institute of Quantum Optics, QUEST-Leibniz Research School, Hanover, Germany.}
\affiliation{Université Paris-Saclay, CNRS, Institut des Sciences Moléculaires d’Orsay, Orsay, France.}

\author{Jannik Ströhle}
\affiliation{Institut für Quantenphysik and Center for Integrated Quantum Science and Technology (IQST), Ulm University, Ulm, Germany.}

\author{Enno Giese}
\affiliation{Technische Universität Darmstadt, Fachbereich Physik, Institut für Angewandte Physik, Darmstadt, Germany.}

\author{Holger Ahlers}
\affiliation{German Aerospace Center (DLR), Institute for Satellite Geodesy and Inertial Sensing, Hanover, Germany.}

\author{Waldemar Herr}
\affiliation{German Aerospace Center (DLR), Institute for Satellite Geodesy and Inertial Sensing, Hanover, Germany.}

\author{Christian Schubert}
\affiliation{German Aerospace Center (DLR), Institute for Satellite Geodesy and Inertial Sensing, Hanover, Germany.}

\author{\'Eric Charron}
\affiliation{Université Paris-Saclay, CNRS, Institut des Sciences Moléculaires d’Orsay, Orsay, France.}

\author{Holger M\"uller}
\affiliation{Department of Physics, University of California, Berkeley, CA, USA.}

\author{Jason R. Williams}
\affiliation{Jet Propulsion Laboratory, California Institute of Technology, Pasadena, CA, USA.}

\author{Ernst M. Rasel}
\affiliation{Leibniz University Hannover, Institute of Quantum Optics, QUEST-Leibniz Research School, Hanover, Germany.}

\author{Wolfgang P. Schleich}
\affiliation{Institut für Quantenphysik and Center for Integrated Quantum Science and Technology (IQST), Ulm University, Ulm, Germany.}
\affiliation{Hagler Institute for Advanced Study, Texas A\&M University, College Station, TX, USA.}
\affiliation{Texas A\&M AgriLife Research, Texas A\&M University, College Station, TX, USA.}
\affiliation{Institute for Quantum Science and Engineering (IQSE), Department of Physics and Astronomy, Texas A\&M University, College Station, TX, USA.}

\author{Naceur Gaaloul}
\affiliation{Leibniz University Hannover, Institute of Quantum Optics, QUEST-Leibniz Research School, Hanover, Germany.}

\author{Nicholas P. Bigelow}
\affiliation{Department of Physics and Astronomy, Institute of Optics, Center for Coherence and Quantum Optics, University of Rochester, Rochester, NY, USA.}

\date{May 29, 2025}

\begin{abstract}
Atom interferometers deployed in space are excellent tools for high precision measurements, navigation, or Earth observation. 
In particular, differential interferometric setups feature common-mode noise suppression and enable reliable measurements in the presence of ambient platform noise. 
Here we report on orbital magnetometry campaigns performed with differential single- and double-loop interferometers in NASA's Cold Atom Lab aboard the International Space Station. 
By comparing measurements with atoms in magnetically sensitive and insensitive states, we have realized atomic magnetometers mapping magnetic field curvatures. Our results pave the way towards precision quantum sensing missions in space.
\end{abstract}

\maketitle

\onecolumngrid

\section{Introduction}

Deploying matter-wave interferometers~\cite{Cronin_2009} to space~\cite{Belenchia_2022,Alonso2022,Abend_2023,NASA2023} promises the advance of fundamental physics by testing general relativity~\cite{Schlippert_2014, Zhou_2015, Asenbaum_2020, Ahlers_2022, He2023}, detecting gravitational waves~\cite{Graham_2013,El_Neaj_2020} and exploring new physics~\cite{Safronova_2018,Tino_2021}.
Their use in applied sensors~\cite{Bongs_2019} could enable, for instance, long-term inertial navigation~\cite{El-Sheimy2020,Gustavson1997,Lenef1997,Li2025}, precising gravity cartography~\cite{Snadden_1998, Rosi_2015, Abend_2016, Hardman_2016} and enhanced magnetic field sensing~\cite{Zhou_2010,Hu_2011,Barrett_2011, Wood_2015, Hardman_2016, Robertson_2017, Hu_2017, Deng_2021}.
For matter-wave interferometers to benefit from space and exceed the performance of classical sensors, they need to exploit extended free-fall times enabling quadratic improvements in sensitivity.
These long interferometry times can best be realized by using slowly expanding Bose-Einstein condensates (BECs)~\cite{Becker_2018,Aveline2020,Deppner2021a,Gaaloul2022} as quantum test masses.

One of the leading systematic effects coupling into the signal of aforementioned fundamental physics experiments~\cite{Asenbaum_2020} are magnetic field gradients.
Fully avoiding such couplings is not possible, neither by operating on a magnetically insensitive sublevel due to the non-linear Zeeman effect, nor by advanced compensation techniques~\cite{McGuirk2002a}.
Thus, magnetic field gradients need to be mapped microscopically at the measurement location, i.e, inside a vacuum chamber.
While magnetic field measurement devices such as superconducting sensors generally offer high precision~\cite{Goodkind1999}, they have to measure from out-of-vacuum, impeding the performance.
In contrast, utilizing atom interferometers themselves would enable precise in-vacuum characterization of magnetic field gradients~\cite{Zhou_2010,Hu_2011,Hardman_2016} at the right spot.
Furthermore, space-based magnetometers are relevant for mapping out kilometer-scale fields, for example, in understanding our Earth's composition~\cite{Olsen2010}, interplanetary missions~\cite{Bennett2021,Amtmann_2024}, navigation~\cite{Muradoglu2025} or climate science~\cite{Courtillot2007}.
Quantum sensors based on cold atoms as test masses are inherently drift-free~\cite{Menoret2018a} and therefore offer additional benefits compared to classical sensors~\cite{Ripka2010}. 

Until now the application of space-deployed BEC interferometers as sensors and at extended free-fall times remains elusive.
Proof-of-principle experiments with BECs were performed using Double Bragg interferometry aboard a sounding rocket~\cite{Lachmann_2021} and Single Bragg interferometry with the Cold Atom Lab (CAL) onboard the International Space Station (ISS)~\cite{Williams2024}.
Additionally, a dual-species BEC interferometer, suitable conceptually for testing general relativity, has been carried out with CAL~\cite{Elliott2023a}.
Finally, using thermal atoms rather than BECs, a Double Raman interferometry based gyroscope was realized in the China Space Station~\cite{Li2025}.
However, so far, all space-borne BEC interferometers were limited in their sensitivity by short interrogation times, mainly due to laser phase and vibrational noise in the non-differential~\cite{Williams2024} or insufficient control of the atom dynamics in the differential case~\cite{Elliott2023a}.
This prevented their use for accurate sensing of external forces or differential gravitational accelerations.

Overcoming previous limitations, we apply a BEC interferometer as a magnetic field sensor in orbit.
The experiments are carried out using CAL, the same setup used for aforementioned sensitivity-limited interferometers~\cite{Elliott2023a,Williams2024}.
To improve the sensitivity, we apply advanced state-engineering techniques~\cite{Corgier_2018,Gaaloul2022} enabling longer spatial overlap of condensed $^{87}$Rb atoms with the interferometer beam.
Additionally, while using Single Bragg pulses, we execute differential interferometry schemes suppressing laser phase noise in the vibrationally noise environment of the ISS.
This allowed the execution of various differential interferometer geometries up to total interrogation times of $2T=\qty{40.3}{\milli\second}$, significantly improving upon previous space-based realizations where interferometer visibilities could not be shown beyond~\cite{Lachmann_2021,Elliott2023a,Williams2024} $2T=\qty{4}{\milli\second}$.

In particular, using condensed $^{87}$Rb atoms in the magnetic sublevel $\mftwo$, we measure a magnetic field curvature inside the ISS payload's vacuum chamber of $|B''| = (614.05 \pm 0.31) \;\si{\nano \tesla \per \square \milli \meter}$ when analyzing all interferometer times globally.
At individual interferometer times $2T=\qtylist{10.3;12.3;16.3}{\milli\second}$, we measure magnetic field curvatures with average uncertainties ranging from $\qtyrange{4.98}{5.06}{\nano \tesla \per \square \milli \meter}$.
Deploying $\mfzero$ atoms, we constrain the acceleration gradient beyond the linear Zeeman effect to $|\Gamma| \le 0.29^{+0.03}_{-0.29} \;\si{\per \square \second}$.
Finally, we obtain differential butterfly measurements with $\mftwo$ atoms, consistent with vanishing third-order magnetic field gradients.

\section{Results}

\subsection{Atom source preparation\label{sec:source_prep}}

\begin{figure}[b]
    \centering
    \includegraphics{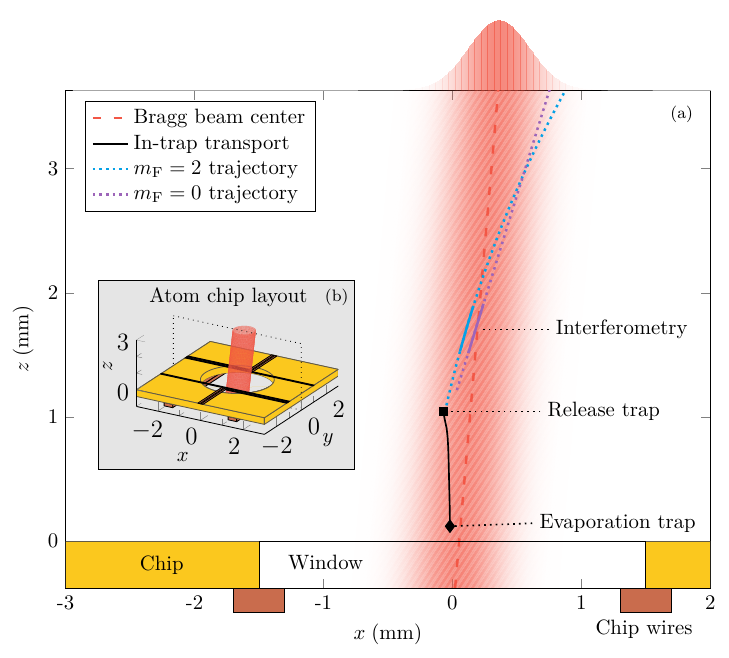}
    \caption{\textbf{Interferometry setup and atom source preparation of the Cold Atom Lab:} (a) Orientation of the Gaussian-shaped Bragg beam (red area) in the $x$-$z$-plane with respect to the atom chip (yellow rectangles) and chip wires (brown rectangles) viewed in the direction of the main imaging axis ($y$-direction). 
    The Bragg beam enters the vacuum chamber through a window of 3 mm diameter in the atom chip under an angle of $4.77 \pm 0.04 ^{\circ}$ with a beam width of $0.47 \pm 0.02$ mm (visualized by the red color gradient) and is retro-reflected by a mirror (not shown). 
    BEC evaporation is performed close to the atom chip in a tight magnetic trap (black diamond) followed by a transport of roughly 1 mm in the $y$ and $z$-directions within 200 ms to the shallower release trap (black square) centered above the chip window. 
    Atoms in the magnetic sensitive $m_\mathrm{F} = 2$ hyperfine state are released with a non-vanishing velocity away from the atom chip and follow a parabolic trajectory (blue dotted line) caused by magnetic forces. Transferring atoms in the magnetic insensitive $m_\mathrm{F} = 0$ hyperfine state by an adiabatic rapid passage 15 ms after release yields an unperturbed linear center-of-mass motion (purple dotted line). 
    Interferometry (see Fig.~\ref{fig:main_figure_2}) is performed in a region where the atoms (solid blue and purple lines) are close to the maximum intensity of the Bragg beam. 
    Here the trajectories during the longest achievable interferometer times $2T = 24.3$ ms and $2T = 40.3$ ms for $\mftwo$ and $\mfzero$ atoms, respectively, are shown. 
    (b) 3D layout of the atom chip and the Bragg beam with the chip window in the center. The dotted rectangle shows the boundaries of subplot (a).}
    \label{fig:main_figure_1}
\end{figure}

The CAL atom chip produces a magnetically trapped BEC of $^{87}$Rb in the $F=2$, $\mftwo$ hyperfine state with approximately $9 \times 10^3$ condensed atoms and a BEC fraction of roughly $70\%$ through evaporative cooling~\cite{Elliott2018, Aveline2020, Williams2024}. 
For atom interferometry, the atoms must be positioned within the Bragg interferometry beam well above the atom chip as illustrated in Fig.~\ref{fig:main_figure_1}.
Thus, after evaporative cooling, we magnetically transport the atoms~\cite{Torrontegui_2011,Corgier_2018, Gaaloul2022} within $\qty{200}{\milli \second}$ over roughly $\qty{1}{\milli\meter}$ in both the $y$- and $z$-direction.
This final trap is aligned with the Bragg beam at a position of $(x,y,z) = (-0.07,\, -0.11,\, 1.04) \; \unit{\milli\meter}$ with trap frequencies $(\omega_x, \omega_y, \omega_z) = 2\pi\, \times (9.28, \,21.2,\, 18.1)\; \unit{\hertz}$.

The CAL interferometry beam has a wavelength of $\qty{785}{\nano \meter}$ and features two adjustable and slightly detuned frequency components building up two standing and two moving optical lattices.
These can be used to couple different momentum states of the $^{87}$Rb atoms via Bragg transitions~\cite{martin1988bragg}.
To ensure resonance with only one of the moving lattices, the atoms need a non-zero momentum $p_0$ along the beam~\cite{Hartmann2020}.
The beam points in $z'$-direction closely aligned with the $z$-axis and tilted by \qty{4.77 \pm 0.04}{\degree} towards the $x$-axis with a beam width of \qty{0.47 \pm 0.02}{\milli\meter} as confirmed by careful calibration measurements of the Bragg beam orientation and intensity (see Methods).
Any transversal velocity of the atoms relative to the beam should be minimized.
We control the atoms' momentum by modifying the trapping potential during a final hold time of \qty{10}{\milli\second} and by briefly applying additional magnetic field gradients along the $x$-axis after release.
In the $\mftwo$ case, this procedure launches the atoms away from the chip with an initial velocity of $v_x = \qty{1.8 \pm 0.1}{\milli \meter \per \second}$ and $v_z = \qty{12.5 \pm 0.2}{\milli \meter \per \second}$.
After this initial procedure, we switch off all chip and coil currents during free propagation of the clouds and obtain a BEC expansion energy $k_\text{B} T_\text{eff} /2$ corresponding to an effective temperature $T_\text{eff}$ of only $\qty{0.66}{\nano\kelvin}$ along the $z$-direction, where $k_\text{B}$ is the Boltzmann constant.
Reaching such small expansion energies is essential to avoid inefficiencies in the Bragg transitions due to Doppler shifts.
This performance is largely enabled by microgravity, allowing the controlled release from such shallow traps and thus avoiding the need for additional collimation techniques.

While the evaporation and transport happens with magnetically sensitive atoms in $\mftwo$, we perform some atom interferometry experiments in the magnetically insensitive state $\mfzero$.
By applying a radio-frequency adiabatic rapid passage~\cite{Williams2024} within the first $\qty{15}{\milli\second}$ after release, we can transfer the atoms into the $\mfzero$ state with high efficiency.
Applying the same adjustable release scheme as in the $\mftwo$ case, we realize an initial velocity of the $\mfzero$ atoms of $v_x = \qty{3.0 \pm 0.1}{\milli \meter \per \second}$ and $v_z = \qty{9.9 \pm 0.1}{\milli \meter \per \second}$.
The trajectories for both hyperfine states are shown in Fig.~\ref{fig:main_figure_1}, clearly indicating that the magnetically sensitive state $\mftwo$ is bent by external forces while the $\mfzero$ atoms follow a straight line.

\subsection{Differential atom interferometry}

\begin{figure}[h]
    \centering
    \includegraphics{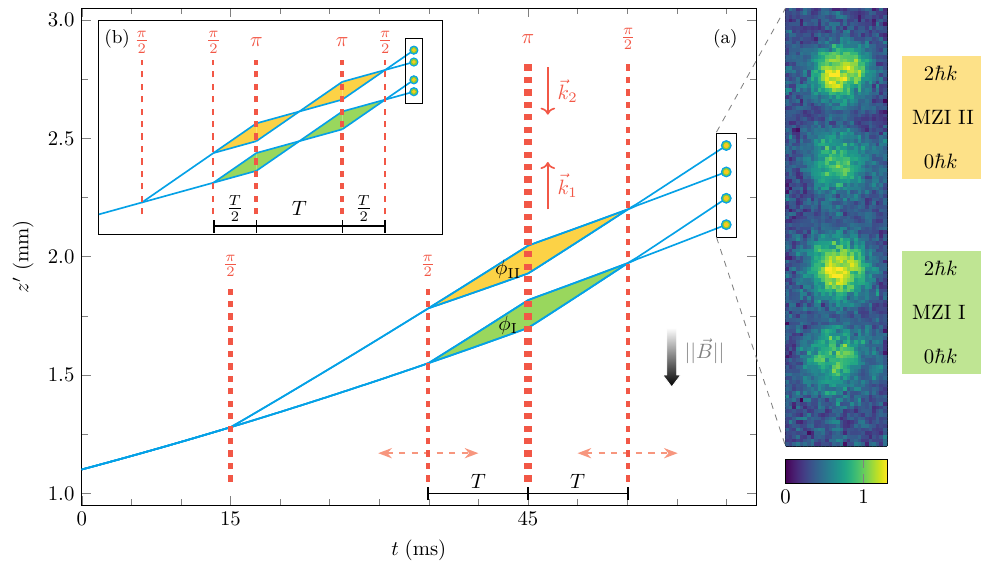}
    \caption{\textbf{Atom interferometry sequence for measuring differential accelerations:} (a-b) Atom trajectories of $^{87}$Rb atoms in the $m_\mathrm{F} = 2$ hyperfine state (blue lines) showing the position during (a) differential Mach-Zehnder-type atom interferometers and (b) differential butterfly interferometers.
    The interferometers consist of four or five laser pulses (red dashed lines), respectively, with effective wave vector $k_\mathrm{eff} = |\vec{k}_1| + |\vec{k}_2| = 2k$ that coherently manipulate the atoms. 
    Positions are measured in the direction $z^\prime$ of the Bragg beam. 
    (a) A first $\pi/2$ splitting pulse is applied 15 ms after release from the trap generating a superposition of atoms in the two momentum states $0\hbar k$ and $2\hbar k$ which each serve as input states for a three-pulse Mach-Zehnder interferometer ($\pi/2$ -- $\pi$ -- $\pi/2$-pulse) that again splits, reflects and recombines the atoms by common laser pulses. 
    After the last laser pulse the four resulting atom clouds spatially separate during 10 to 15 ms of free evolution and are detected by a single absorption image (see exemplary density plot for $2T = 20.3$ ms, size $117 \times 488$ \textmu m$^2$). 
    The total interferometer time $2T$ is varied while the $\pi$-pulse is always applied 30 ms after the initial splitting pulse. 
    Accelerations of the atoms with respect to the laser beams, caused for instance by spatially-dependent magnetic fields $\vec{B}(z^\prime)$, lead to non-vanishing interferometer phases $\phi_\mathrm{I}$ and $\phi_\mathrm{II}$ affecting the relative population in the output ports and enable measurement of differential accelerations between the two MZIs.
    (b) By adding a second $\pi$ mirror pulse during the interferometer sequence a differential figure-eight or butterfly geometry can be realized which in leading order is insensitive to acceleration gradients, but instead reveals acceleration curvatures if present.
    }
    \label{fig:main_figure_2}
\end{figure}

We perform two types of differential atom interferometry geometries to characterize the gradient and curvature of the local force field acting on the atoms.
For characterization of the force gradient, we employ a differential Mach-Zehnder type geometry defined by the four-pulse sequence shown in Fig.~\ref{fig:main_figure_2}a, and apply it separately on the magnetic sublevels $\mftwo$ and $\mfzero$.
To characterize the force curvature, we extend this to a double-loop differential butterfly geometry by adding a second mirror pulse as indicated in Fig.~\ref{fig:main_figure_2}b.

To generate the differential geometry, we create two spatially separated Mach-Zehnder interferometers (MZI) and apply all beam splitter and mirror pulses simultaneously on both atomic clouds.
We establish their spatial separation by performing a $\pi/2$ Bragg transition on the atoms with initial momentum $p_0$ to create an equal superposition of momentum states $p_0$ ($\zerohk$) and $p_0+\hbar k_\text{eff}$ ($\twohk$).
Here, the effective wave vector $k_\text{eff}$ is given by the Bragg beam's resonant moving lattice, where $k_\text{eff} = | \vec{k_1} - \vec{k_2}|$ while $|\vec{k_i}| = 2\pi / \lambda_i$ describes the wave vector of two slightly detuned frequency components.
The pulse duration is $\qty{0.13}{\milli\second}$.
During some time of flight the two momentum states $\zerohk$ and $\twohk$ spatially separate and serve as the input states for \mbox{MZI-$\mathrm{I}$} and MZI-$\mathrm{II}$, respectively.
We have analyzed the differential center-of-mass (COM) motion between these two momentum states for expansion times up to \qty{150}{\milli\second} and extracted an estimate of the acceleration gradient in $z'$-direction of $|\Gamma| = 36.76 \pm 1.07 \;\si{\per \square \second}$ experienced by the $\mftwo$ atoms in the system (see Methods and Discussion).

Each of the two MZIs illustrated in Fig.~\ref{fig:main_figure_2}a accumulates, to leading order, a phase $\phi_j = k_\text{eff}a_jT^2$ with $j = \mathrm{I}, \mathrm{II}$ referring to the individual MZIs, the common pulse separation time $T$ measured from the center of the pulses and the local acceleration $a_j$ acting on the atoms along the direction of the Bragg beam.
Measuring these phases individually suffers from laser phase and vibrational noise and becomes infeasible for extended pulse separation times $T$ in noisy environments such as CAL~\cite{Williams2024}.
However, by extracting the phase difference $\Delta\phi = \phi_\mathrm{II} - \phi_\mathrm{I}$, common phase-noise contributions cancel out~\cite{Fang_2016, Struckmann_2024} and the evaluation of the differential MZI becomes independent of this constraint.
Consequently, from the differential MZI phase $\Delta \phi$, the local acceleration gradient $\Gamma$ which is assumed to be constant over the experiment region can be evaluated according to the analytic phase relation $\Delta\phi = k_\text{eff} \Gamma \Delta z' T^2$, where $\Delta z'$ is the distance between the individual MZIs.
In our case, $\Delta z' = 2 v_\text{rec} t_\text{sep}$ with the recoil velocity $v_\text{rec} = \qty{5.85}{\milli\meter \per \square\second}$ and the separation time $t_\text{sep} = \left(\qty{30}{\milli\second} - T\right)$ between the two momentum states prior to the three-pulse MZI sequence. 
The measurement of $\Delta \phi$ is based on determining the respective atom numbers in all four spatially resolved exit ports from a single experimental shot, an example shown in Fig.~\ref{fig:main_figure_2}.
Comparing the relative populations \mbox{$N_{\mathrm{rel},j} = N_{0\hbar k, j}/(N_{0\hbar k, j} + N_{2\hbar k, j})$} of the $\zerohk$ states in both MZIs yields correlation plots with varying shape depending on the differential phase $\Delta\phi$.

\begin{figure}[h]
    \centering
    \includegraphics{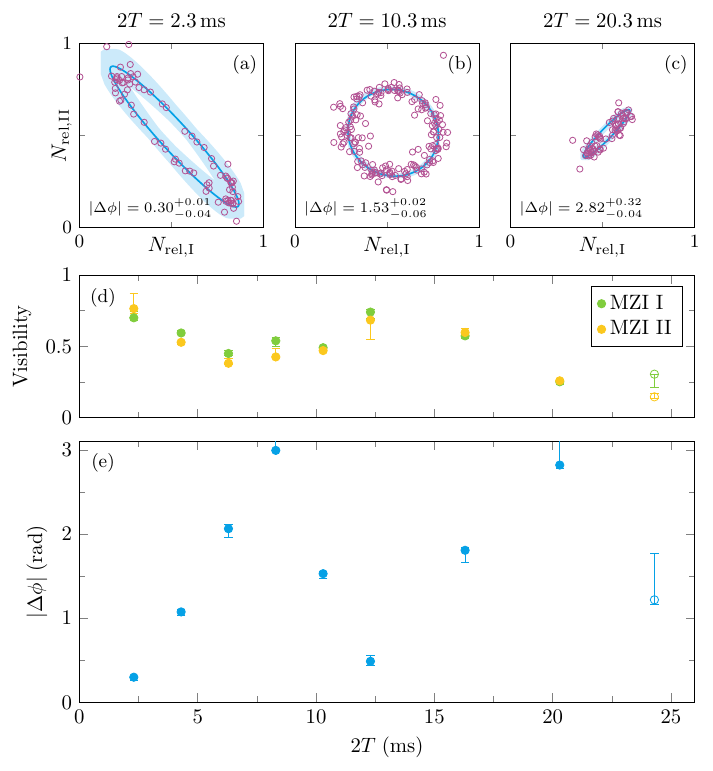}
    \caption{\textbf{Experimental results of differential Mach-Zehnder atom interferometry in space:} 
    (a-c) Experimental data of the relative populations $N_\mathrm{rel,I}$ and $N_\mathrm{rel,II}$ in the $0\hbar k$ state of both Mach-Zehnder interferometers MZI I and MZI II described in Fig.~\ref{fig:main_figure_2}~(a) with atoms in the magnetically sensitive $\mftwo$ state. 
    For growing interferometer time $2T$ the shape of the correlation data (purple circles) varies between elongated ellipses and circles indicating a changing differential phase $\Delta\phi = \phi_\mathrm{II} - \phi_\mathrm{I}$ obtained by ellipse fitting (blue lines with confidence areas in shaded blue) of the data (see Methods).
    (d) Visibility of each MZI determined by the maximum extend of the ellipse fitted to the correlation data with error bars showing the 1$\sigma$ ellipse-fitting confidence bounds.
    Altering visibility is due to day-to-day performance of the experimental apparatus with respect to atom numbers and BEC fraction and in general decreases for long interferometer times due to motion and expansion of the atoms perpendicular to the Bragg beam (see Fig.~\ref{fig:main_figure_1}).
    The data at $2T=\qty{24.3}{\milli\second}$ is not considered for further analysis due to the poor visibility of less than $15\%$ in MZI II and resulting large phase uncertainty.
    (e) Measured differential phase for each interferometer time $2T$ illustrating the presence of differential accelerations. Error bars are based on 1$\sigma$-confidence bounds of the bootstrap analysis. For $2T = 8.3$ ms and $2T = 20.3$ ms the error bars were extended to include $\pi$ since in these cases the data did not allow a clear differentiation from this value (see Methods).
    }
    \label{fig:main_figure_3}
\end{figure}

We perform a differential MZI measurement campaign with $\mftwo$ atoms.
To this end, we employ nine different interferometer times $2T$ ranging from $\qtyrange{2.3}{24.3}{\milli\second}$ with an average number of $96$ experimental shots per interferometer time.
Distributed over these shots, we scan the laser phase of the last Bragg pulse from $0$ to $2\pi$ in steps of $\pi/6$ in order to realize different populations in the respective exit ports.
Due to laser phase noise, this discrete manual phase scan becomes blurred and in principle even unnecessary for most of the used interferometer times.
During the interferometer sequence, the atoms experience an acceleration along the beam direction.
To ensure resonance with the mean velocity of the atom clouds, we adjust the Bragg pulse two-photon detunings for each interferometer time and each pulse, individually.
The three pulse durations are constant over all sequences at \qtylist[list-units = bracket]{0.1; 0.2; 0.1}{\milli\second}, respectively.
In Fig.~\ref{fig:main_figure_3}a-c, we exemplary show correlation plots as the measurement outcome for three different interferometer times from this campaign.

We determine $\Delta\phi$ by fitting a rotated and shifted ellipse to the correlation data and evaluate its uncertainty by a bootstrapping method.
This bootstrapping method estimates the statistical distribution of the differential phase by generating multiple datasets from the experimental data via sampling with replacement.
By fitting each bootstrap dataset individually, we calculate \mbox{1$\sigma$-confidence} bounds for all variables.
Due to the different initial momentum states of the two MZIs, a constant phase offset of $\pi$ emerges in the measurements which we have subtracted in all displayed differential phase values $\Delta\phi$ since it is merely a choice of exit port labeling. 
For strongly elongated and noisy correlation data, i.\,e., close to degeneracy, it is inherently hard for ellipse fitting techniques to distinguish between truly non-zero differential phase values and vanishing differential phases~\cite{Foster2002a,Stockton_2007,Rosi_2015,Zhang2023}.
The same holds true for values very close to $\pi$. The data shown in Fig.~\ref{fig:main_figure_3}a-c clearly illustrate this point. On the one hand in Fig.~\ref{fig:main_figure_3}a the points form a slightly opened ellipse with no points populating its center, yielding a truly non-zero differential phase. On the other hand in Fig.~\ref{fig:main_figure_3}c this assessment cannot be done reliably. 
Therefore, in these cases we extend the error bars to include $0$ or $\pi$, respectively (refer to Methods for more details on the ellipse fitting technique).

The main limitation for extracting the differential phase for long interferometer times stems from the vanishing visibility of population oscillations in the individual MZIs.
This is caused by the atom's COM motion relative to the Bragg beam and their transverse expansion such that the atom cloud explores regions of reduced Bragg beam intensity.
These effects lead to a decreased transfer efficiency for long interferometer times and therefore pose a limit for the maximal time $2T$ that could be performed in the $\mftwo$ case.
We determine the individual visibilities from the maximum vertical and horizontal extent of the fitted ellipses, respectively, shown in Fig.~\ref{fig:main_figure_3}d.
The visibility decreases for growing interferometer times until their small values significantly impede the differential phase determination, reflected in increasing uncertainties. 
For the $2T = \qty{24.3}{\milli\second}$ campaign the visibility of MZI-II drops below $0.15$ rendering the phase estimation unreliable such that we are not considering the outcome of this campaign for further quantitative analysis.

The results of the differential phase fits are displayed in Fig.~\ref{fig:main_figure_3}e for all interferometer times $2T$.
Due to multiple ambiguities in inferring the differential phase from ellipse fits, all measurements are limited to absolute phases in between $0$ and $\pi$.
This ambiguity prevents the direct determination of the acceleration gradient from individual interferometer times alone.
However, by relating the differential phases from all interferometer times, we can evaluate the acceleration gradient.
For small interferometer times $2T \le \qty{6.3}{\milli\second}$ the measured differential phase values slowly increase and can be connected by the analytic phase relation for $\Delta\phi$ to obtain a short-time estimate of $\Gamma$. 
By extending this relation to larger times, we resolve the phase ambiguity for all measurements with $2T\ge \qty{8.3}{\milli\second}$ by choosing the best match step by step.
Figure~\ref{fig:main_figure_4} shows the resulting differential phases after resolving all ambiguities.
The acceleration gradient from this global fit over all interferometer times is then given by $|\Gamma| = \qty{39.46 \pm 0.02}{\per \square\second}$.

We also evaluate the force gradient on magnetically insensitive $\mfzero$ atoms.
In contrast to the magnetically sensitive state, the $\mfzero$ atoms are not significantly accelerated by residual magnetic field gradients during the interferometer sequence, reducing their transversal motion relative to the Bragg beam.
This improves the Bragg transition efficiencies for long times of flight, leading to better visibilities for long interferometer times.
The approach allows to extend the interferometer time up to $2T = \qty{40.3}{\milli\second}$ while still observing a signal at the exit ports.
The measured differential phases are shown in Fig.~\ref{fig:main_figure_4} with no clear dependence on the interferometer time.
Based on the fact that all phase values lie well in the small subregion $[0,\pi/6]$, we rule out that a relation to the interferometer time is merely hidden in the ellipse fit phase ambiguity $\phi_\text{meas} \in [0,\pi]$.
Moreover, for all $\mfzero$ campaigns the correlation data does not allow a clear distinction whether the differential phases are truly non-zero or not and, consequently, the error bars include $\Delta\phi = 0$. 
Thus, for the $\mfzero$ atoms, we evaluate an upper bound of the acceleration gradient based on only the longest interferometer time $2T=\qty{40.3}{\milli\second}$ yielding a value of $|\Gamma| \le 0.29^{+0.03}_{-0.29} \;\si{\per \square \second}$.

We further extend the differential MZI configuration to a differential butterfly interferometer (BFI) to measure force curvatures instead of gradients.
The differential BFI consists of two spatially separated butterfly configurations~\cite{Canuel2006} as shown in Fig.~\ref{fig:main_figure_2}b.
Compared to the differential MZI, it features an additional mirror pulse before the closing beam splitter.
The resulting symmetry eliminates the leading term that dominated the differential MZI signal, such that it reveals smaller contributions due to force curvatures (see Methods).
We perform a measurement campaign with the differential BFI using $\mftwo$ atoms.
For three different interferometer times $2T$ ranging from $\qtyrange{4.5}{20.5}{\milli\second}$, we again see no clear relation between the differential phase and the interferometer time, as shown in Fig.~\ref{fig:main_figure_4}.
Noise in the relative port population makes an extraction of $\Delta\phi$ close to degeneracy of the ellipse challenging, but the obtained results are compatible with zero force curvature $\Gamma_\text{I}-\Gamma_\text{II} \approx 0$.
Although anharmonic potentials also affect the differential MZI signal, our butterfly measurements indicate that accounting solely for the leading-order terms above, that is, the magnetic field curvature, is sufficient for analyzing the differential MZIs.

\subsection{Magnetometry}

Combining the results from all measurement campaigns, we locally characterized the magnetic field.
The acceleration gradient observed in the differential MZI with magnetically sensitive $\mftwo$ atoms almost fully vanishes for magnetically insensitive $\mfzero$ atoms, highlighting its magnetic nature.
We thus conclude the presence of a local magnetic field curvature along $z'$ of \mbox{$\left|B''\right| =\left|\Gamma \cdot m_\text{Rb} / \left( m_F g_F \mu_B\right)\right|= \qty{614.05 \pm 0.31}{\nano\tesla \per \square \milli \metre}$} with less than $0.8\%$ of contributions outside of the linear Zeeman effect quantified through the upper bound of $\Gamma$ acting on $\mfzero$ atoms.
Furthermore, our differential BFI measurement is consistent with the third derivative of the magnetic field being zero.

Instead of inferring the differential acceleration by fitting $\Delta \phi$ as a function of $T$ for the entire data set, we can also calculate it directly for each value of $T$, analyzing each interferometer time individually.
The global analysis resolved the ambiguity in the differential phases $\Delta \phi ^\text{MZI}$ measured with the $\mftwo$ atoms.
Using the same phase relation as before, the acceleration gradient $\Gamma$ and thus the magnetic field curvature $B''$ is determined via the linear Zeeman effect.
Fig.~\ref{fig:main_figure_5} shows the resulting curvatures for individual interferometer times.
The most sensitive measurements are realized for interferometer times ranging from \mbox{$2T=\qtyrange{10.3}{16.3}{\milli\second}$}, featuring magnetic field curvatures with average uncertainties between $\qtylist{4.98;5.06}{\nano \tesla \per \square \milli \meter}$.
For example, we measure $\left|B''\right| = 599.27^{+7.16}_{-2.8} \;\si{\nano \tesla \per \square \milli \meter}$ using $2T=\qty{10.3}{\milli\second}$.
This experiment includes $181$ shots with an average atom number of $4.5 \times 10^3$ atoms per MZI and visibility of $0.48$.
Based on these numbers, the corresponding 1$\sigma$-shot-noise limit for the magnetic field curvature amounts to $\qty{0.4}{\nano \tesla \per \square \milli \meter}$.
The gap between the experimentally achieved sensitivity and the theoretical limit can have a variety of causes. In particular, we suspect signal and experiment drifts in between shots, noisy imaging at small atom numbers, and consequently non-ideal ellipse fitting performance.

In Fig.~\ref{fig:main_figure_5} we observe a spread of the individually analyzed data around the globally fitted result and between each other.
This results in a weighted average for the individual interferometer times of $\left|B''\right| = \left(622.08 \pm 19.66 \right) \si{\nano \tesla \per \square \milli \meter}$ with its standard deviation being larger than most individual uncertainties. 
To understand this spread we have performed a detailed numerical modeling taking into account more experimental parameters such as finite pulse durations~\cite{Antoine_2006} as well as additional phase contributions within a quadratic potential~\cite{Antoine_2003,Roura_2014,Roura_2017,Roura_2020}.
Comparing these numerical results to the ones from our analytical phase relation reveals only a small offset on the order of few \qty{}{\nano\tesla \per \square \milli\meter} while not decreasing the observed spread, confirming the validity of our simpler analytical phase model.
Furthermore, higher-order contributions beyond a quadratic potential were constrained by our differential BFI measurements already, and including these in our phase model does not yield a consistent and improved explanation of the data (see Methods).
As the experiments were performed over a duration of several months, we conclude that actual drifts in the magnetic field curvature are the most likely explanation for the observed data spread.
While with CAL as a multi-user facility, such long campaign durations are hard to avoid, this does not constitute a general limitation of the here presented methods when deployed in dedicated magnetometry missions.

\section{Discussion}

In this article we have performed various differential atom interferometer sequences based on Mach-Zehnder and butterfly geometries to characterize the magnetic field inside the CAL vacuum chamber onboard the ISS. 
In particular, we have measured the residual magnetic field curvature with a 1$\sigma$-sensitivity of $\qty{0.31}{\nano \tesla \per \milli \meter \squared}$ and have obtained complementary measurements with magnetically insensitive atoms to confirm the magnetic nature of the measured force gradients. 
Furthermore, we verified through the differential butterfly interferometer that there is no significant contribution from a third spatial derivative of the magnetic field acting on the atoms and the measured differential forces can be attributed to a magnetic field curvature.
With these results we have successfully realized the first BEC-based magnetometer in space representing an important milestone for space-based quantum sensors in general.

The residual magnetic fields present within the CAL device have been observed earlier by studying the COM motion of the atoms~\cite{Pollard_2020, Gaaloul2022} and can best be explained by magnetic fields originating from the atom-chip connectors or other magnetized parts of the experimental setup. 
Their precise characterization was only possible by applying interferometry techniques as presented in this work. 
The measured magnetic field curvature corresponds to an effective harmonic trap frequency of $\omega_{z'} = 2\pi \times \qty{1.000 \pm 0.002}{\hertz}$ along the direction of the Bragg beam and thus, would influence both the atom's COM motion as well as their expansion dynamics when employed for longer times required for precision sensing campaigns.

Comparing the magnetic field curvature result of $\left|B''\right| = \qty{614.05 \pm 0.31}{\nano\tesla \per \square \milli \metre}$ from our differential interferometer to $\left|B''\right| = \qty{572 \pm 17}{\nano\tesla \per \square \milli \metre}$ obtained by a differential COM measurement with atoms in the $\zerohk$ and $\twohk$ state illustrates the higher sensitivity of the interferometer.
In addition, the independent result from the COM motion confirms that the process of resolving the phase ambiguities for each interferometer time was done correctly, as both results are close together.
The existing offset between both approaches can be attributed to the fact that the interferometer measures much more locally than the classical measurement.
The interferometer takes place only during a small fraction of the total atom cloud trajectory (\qty{0.43}{\milli\meter} motion during a $2T = \qty{20.3}{\milli\second}$ interferometer compared to \qty{3.5}{\milli\meter} during \qty{130}{\milli\second} differential time-of-flight) and thus introduces less averaging.

Consequently, our quantum sensor is not just relevant in its own right, but also verifies the magnetic field environment of a space-deployed atom interferometer. 
This careful verification is mandatory for any high-precision gravity measurement to estimate the unwanted phase shifts from residual magnetic fields~\cite{Hu_2017, Deng_2021}. 
In most cases the magnetic field environment is mapped out by Ramsey interferometer techniques that allow for a determination of the magnetic field gradient in a certain spatial region~\cite{Wood_2015,Deng_2021}. 
In contrast, our approach also enables a direct measurement of the higher derivatives of the magnetic field and thus provides deeper insights into the local magnetic field environment of the apparatus. 
Furthermore, our differential interferometer can operate in a much smaller volume compared to Earth-bound measurements and thus improves the spatial resolution of the magnetic field characterization.

In order to achieve the results presented in this article, a detailed characterization of the Bragg beam orientation and intensity as well as the atom source was crucial to enable the required spatial overlap between the atom cloud and the Bragg beam during the interferometer. 
In this way we have substantially improved the single mirror pulse efficiencies up to \qty{85}{\percent} outperforming previous atom interferometer implementations of CAL~\cite{Elliott2023a, Williams2024} .
This Bragg beam characterization constitutes another Bragg-pulse enabled quantum sensor used for validation of the experimental setup by itself. 

In addition, the microgravity conditions aboard the ISS enabled our measurements in the following ways:
First, a freely falling laboratory allows to probe the atoms for long times at a constant position or close to a specific object, facilitating characterization of the system and sensing applications. 
In our case the atoms moved by only \qty{0.43}{\milli\meter} during each MZI for an interferometer time of $2T = \qty{20.3}{\milli\second}$ and thus enabled a very localized determination of the magnetic field curvature. 
Performing the same measurement on Earth would result in a total motion of \qty{9.0}{\milli\meter} between the beam splitting pulses of the MZI such that the measurement signal would be averaged over a longer distance potentially reducing the sensitivity.  
Second, microgravity allows to utilize shallow release traps to reduce the expansion energy of the BEC and to enable compact atom clouds even after long free evolution times. 
By carefully designing the transport of the BEC to the release trap and optimizing the release scheme we achieved BEC expansion energies of \qty{0.66}{\nano\kelvin} which are much lower compared with the proof-of-principle interferometry demonstrations on CAL.  Consequently, we were able to increase the pulse efficiency of the Bragg transitions and thus extend the interferometer times to $2T = \qty{40.3}{\milli\second}$. 
Applying delta-kick collimation techniques (DKC)~\cite{Ammann_1997,Deppner2021a,Gaaloul2022,Albers_2022,Herbst_2024} to further reduce the expansion energy of the BEC was unnecessary in our case, as the achieved expansion rates were sufficiently low and did not constitute the limiting factor during our campaign.

In contrast, the total number of condensed atoms was a true limitation of the current implementation. 
Routinely having an average of roughly $2 \times 10^3$ atoms per exit port renders the atom number determination through absorption imaging challenging especially when interference reduces the actual atom number well below $10^3$ in single exit ports. 
These low numbers introduce noise in the correlation data and increase the uncertainty of phase estimation. 
Increasing the total number of condensed atoms from around $10^4$ to $10^6$, as for instance envisioned for the BECCAL apparatus~\cite{Frye_2021}, would improve the signal-to-noise ratio considerably.
Furthermore, the Bragg beam properties and alignment with the atom trajectories could be improved to preserve the contrast of the entire interferometer for longer times. 
In particular, a larger Bragg beam diameter would minimize the effects of the whole atom cloud moving transversal to the beam due to magnetic forces and reduce losses during the pulses.
Similarly, a higher laser power would enable much shorter beam splitter and mirror pulses increasing their efficiencies.
Lastly, the Bragg beam experiences diffraction effects from the edge of the chip window and the chip wires compromising the uniformity of the beam. 
Adjusting the beam path could reduce these unwanted effects and further improve the pulse efficiencies. 
These aspects are the main reasons why our implementation did not reach the sensitivities of dedicated ground-based differential atom interferometry setups~\cite{Rosi_2015, Werner2024a}, that are less limited by space, weight and power constraints. 
However, an upgraded facility could overcome these challenges and reach unprecedented sensitivities in the future~\cite{Struckmann_2024}.

In summary, we have for the first time employed a differential atom interferometer to map out the magnetic field environment of a space-based BEC machine. 
This constitutes a major step forward for quantum sensing in space and also paves the way for future Earth observation missions utilizing differential atom interferometry. 
In fact, the methods applied here match those required for gravity gradient sensing based on cold atom technology and therefore are a true pathfinder for more sophisticated space missions~\cite{Alonso2022, Abend_2023}.

\begin{figure}[h]
    \centering
    \includegraphics{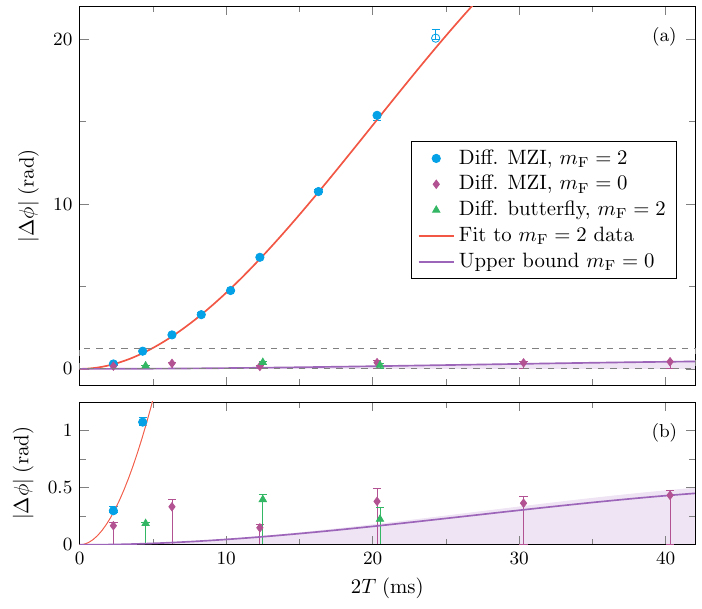}
    \caption{\textbf{Comparison of complementary differential interferometers:} 
    (a) Measured differential phase $\Delta\phi$ as a function of the interferometer time $2T$ for the differential MZI sequence shown in Fig.~\ref{fig:main_figure_2}a performed with atoms either in the $m_\mathrm{F} = 2$ (blue dots) or $m_\mathrm{F} = 0$ (purple diamonds) hyperfine state as well as results for $m_\mathrm{F} = 2$ atoms in a differential butterfly geometry (green triangles), featuring two $\pi$-pulses instead of one (see Fig.~\ref{fig:main_figure_2}b).
    Error bars are obtained by ellipse fitting uncertainty (see Methods). None of the data sets for the $m_\mathrm{F} = 0$ and the butterfly interferometer allow a clear differentiation from $\Delta\phi = 0$, such that for these campaigns the error bars were extended to include zero.
    (b) Magnification of (a) for small values of $\Delta\phi$ illustrating that for $m_\mathrm{F} = 0$ atoms and for the butterfly interferometer all measured differential phases are quite small and do not show a clear increasing behavior. 
    A phase relation fit (red) to the $m_\mathrm{F} = 2$ data (blue) reveals an acceleration gradient $|\Gamma| = \qty{39.46 \pm 0.02}{\per \square\second}$.
    In contrast the $m_\mathrm{F} = 0$ data (purple) shows an upper bound for a non-magnetic acceleration gradient based on the longest interferometer time $2T=\qty{40.3}{\milli\second}$ of $|\Gamma| \le 0.29^{+0.03}_{-0.29} \;\si{\per \square \second}$.
    The butterfly results are compatible with zero force curvature.
    Overall, the data show clear evidence of a differential magnetic force acting on the atoms.
    }
    \label{fig:main_figure_4}
\end{figure}

\begin{figure}[h]
    \centering
    \includegraphics{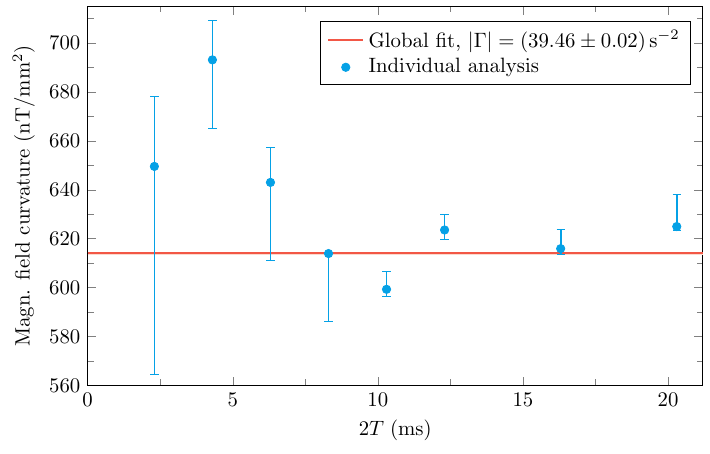}
    \caption{
    \textbf{Measured magnetic field curvature:}
    Using the data from the differential MZI measurement performed on $\mftwo$ atoms, for each interferometer time $2T$, we individually infer the magnetic field curvature \mbox{$\left|B''\right| =\left|\Gamma \cdot m_\text{Rb} / \left( m_F g_F \mu_B\right)\right|$}.
    The global fit shown in Fig.~\ref{fig:main_figure_4} gives $\left|B''\right| = \left(614.05 \pm 0.31 \right) \si{\nano \tesla \per \square \milli \meter}$.
    The weighted average for the individual measurements is $\left|B''\right| = \left(622.08 \pm 19.66 \right) \si{\nano \tesla \per \square \milli \meter}$.
    The best individual performances are obtained for interferometer times ranging from \mbox{$2T=\qtyrange{10.3}{16.3}{\milli\second}$} including 65 to 181 shots, respectively, resulting in average uncertainties in the magnetic field curvatures between $\qtylist{4.98;5.06}{\nano \tesla \per \square \milli \meter}$.
    }
    \label{fig:main_figure_5}
\end{figure}

\clearpage

\section*{Methods}

\subsection*{Characterization of Bragg transitions and beam orientation}
We manipulate the atom's momentum states with a Bragg laser beam at $\qty{785}{\nano\meter}$ operated with box-shaped pulses.
The maximum laser output power of $\qty{66}{\milli \watt}$ is split into two frequency components with adjustable wave vectors $\vec{k}_1$ and $\vec{k}_2$ where $|\vec{k_i}| = 2\pi / \lambda_i$.
By controlling the two-photon detuning $\Delta$ between the two laser frequency components, we enable resonant Bragg transitions only between an initial momentum $p_0$ ($\zerohk$ state) and the momentum $p_0 + \twohk$ ($\twohk$ state).
Here we take into account both the recoil effect on the atoms and the Doppler shift to identify the resonant detuning, such that transitions to other momentum states are well suppressed~\cite{Hartmann2020}.

The transfer efficiency between the two momentum states $\zerohk$ and $\twohk$ strongly depends on the local Bragg beam intensity. 
It is therefore particularly important to carefully characterize the Bragg beam to optimize the diffraction efficiency at different positions.
As CAL is fully remote controlled and encapsulated during operation, it is not possible to measure the beam orientation or intensity from the outside and to confirm its specifications after installation onboard the ISS. 
Taking advantage of the adjustable release scheme described in the results section, we realized different trajectories for atoms in the $\mfzero$ and $\mftwo$ states.
The trajectory of the atom cloud was determined by recording the center-of-mass position of the atoms for different time of flights using both camera systems.
For each trajectory and center-of-mass velocity, we determined the optimal detuning $\Delta$ via individual resonance scans yielding values ranging from $\qtyrange{40}{55}{\kilo\hertz}$.
In each case, we applied a single Bragg pulse with varying duration from $\qtyrange{0.1}{1.6}{\milli\second}$ to drive Rabi oscillations, where the lower limit of $\qty{0.1}{\milli\second}$ was set by technical constraints.
By measuring the populations in the $\zerohk$ and $\twohk$ state after the Bragg pulse and taking into account the BEC expansion rate and pulse shape, we obtain the local Rabi frequency of the laser beam at the respective BEC positions during the pulse.

\begin{figure}[h]
    \centering
    \includegraphics[scale=1]{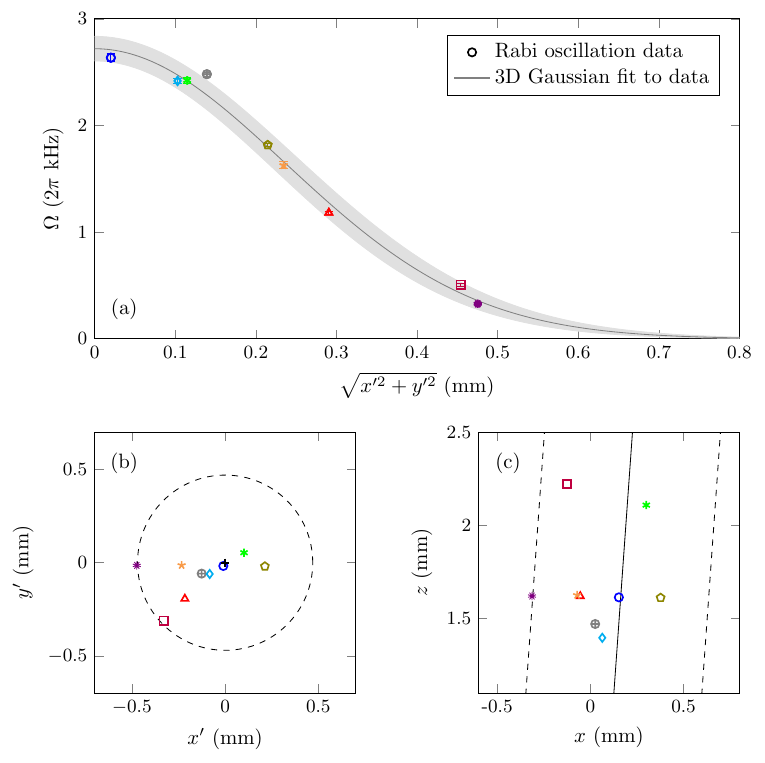}
     \caption{\textbf{Validation of Bragg beam orientation and intensity:} (a) Rabi frequency $\Omega$ of the Bragg beam as a function of the distance from the center of the beam $r^\prime = \sqrt{(x^\prime)^2 + (y^\prime)^2}$ measured by varying the duration of a single Bragg pulse with atoms at different positions. The error bars correspond to the fit uncertainty of the individual Rabi oscillations (see Methods). 
     A three-dimensional Gaussian intensity profile (gray line and gray shaded area for confidence bounds) is fitted to the measured data points (colored markers) yielding a maximum intensity of the Bragg beam or Rabi frequency of $\Omega_\mathrm{max} = 2\pi\,(2.7 \pm 0.1)$ kHz and a beam width of $w = 0.47 \pm 0.02$ mm. 
     (b) Spatial distribution of the individual intensity measurements in the $x^\prime$-$y^\prime$-plane corresponding to a view in the direction of the Bragg beam. The origin of the coordinate system marks the center of the beam and the dashed black circle indicates where $r^\prime = w$.
     (c) Positions of the intensity measurements in the $x$-$z$-plane corresponding to the view of Fig.~\ref{fig:main_figure_1}~(a), where the center of the beam is illustrated by the black solid line and the dashed black lines show where the distance from the center is equal to the beam width. 
     The angle of the Bragg beam with respect to the $z$-axis was determined to $4.1 \pm 3.9 \degree$ with this campaign. 
     The color and marker style is unique for each measurement point and identical in all three plots. 
     }
    \label{fig:supplementary_figure_A1}
\end{figure}

Since the Rabi frequency is proportional to the intensity of the Bragg laser which features approximately a Gaussian intensity profile, the local Rabi frequency depends on the transverse position and we expect the form
\begin{align}
    \Omega(x', y') = \Omega_\text{max} \,\mathrm{e}^{-2[(x' - x'_0)^2 + (y' - y'_0)^2]/w_0^2} \label{eq:method:intensity_beam}
\end{align}
with the maximum Rabi frequency $\Omega_\text{max}$ as well as the width $w_0$, and the centers $x'_0$ and $y'_0$ in $x'$- and $y'$-direction of the Bragg beam.
Since there is a small angle $\alpha$ between the Bragg beam and the $z$-axis in the $x$-$z$-plane, the coordinates transform in the following way
\begin{subequations}
\label{eq:method:coordinate_trafo}
\begin{align}
    x' &= x\cdot\cos\alpha + z\cdot\sin\alpha \label{eq:method:x_coordinate_trafo} \\
    y' &= y \label{eq:method:y_coordinate_trafo} \\
    z' &= -x\cdot\sin\alpha + z\cdot\cos\alpha \,. \label{eq:method:z_coordinate_trafo} 
\end{align}
\end{subequations}
By fitting the model function, Eq.~\eqref{eq:method:intensity_beam}, together with the coordinate transformations, Eq.~\eqref{eq:method:coordinate_trafo}, to the Rabi scan data, we obtain the maximum Rabi frequency $\Omega_\text{max} = 2\pi\cdot \qty{2.72 \pm 0.12}{\kilo\hertz}$ and the following geometric parameters:
the beam width $w_0 = \qty{0.47 \pm 0.02}{\milli\meter}$, the beam center $x_0 = \qty{48 \pm 135}{\micro\meter}$ and $y_0 = \qty{92 \pm 33}{\micro\meter}$ at $z=0$ as well as the angle in the $x$-$z$-plane $\alpha = \qty{4.08 \pm 3.87}{\degree}$.
Here the relatively large error for $\alpha$ is due to the fact that most measurements were performed at similar $z$-positions to better sample the transverse shape of the Bragg beam.
The intensity profile of the Bragg beam and the positions of the individual Rabi scan measurements are displayed in Fig.~\ref{fig:supplementary_figure_A1}.

We have achieved maximum transfer efficiencies of $85\%$ for single mirror pulses performed close to the beam center showing an improvement compared to previous pulse efficiencies demonstrated with CAL \cite{Elliott2023a}.
This result is in good agreement with the maximum theoretical mirror pulse efficiency of $86\%$ determined by the measured Bragg beam intensity and the momentum width of the atom cloud.

A complementary approach to measure the angle of the Bragg beam more precisely relies on the comparison of the COM trajectories of the $\zerohk$ and $\twohk$ state in a single experimental run. 
The distance of the atom clouds in these two states in the $x$-$z$-plane is plotted in Fig.~\ref{fig:supplementary_figure_A2} for expansion times up to \qty{150}{\milli\second} enabling a very sensitive determination of the Bragg beam angle of $\qty{4.77 \pm 0.04}{\degree}$.
By additionally observing the atoms in the $x$-$y$-plane we observe no significant momentum transfer by the Bragg beam in the $y$-direction as expected.

As a last characterization, we employed the differential COM motion of the $\zerohk$ and $\twohk$ states to verify the camera magnification in-flight. 
The total transferred velocity $\twohk/m = \qty{11.70}{\milli\meter\per\second}$ is well defined and can be compared with the observed differential velocity measured with an uncertainty of only $\qty{0.02}{\milli\meter\per\second}$. 
This measurement suggests a correction of $6.2 \pm 0.2\%$ with respect to the original calibration of the camera system. 
Notably, when comparing the results of a differential interferometer for such changes of the magnification, there is only a negligible influence on the relative atom numbers in the exit ports and the measurement data extracted from them. 
Hence, our interferometric results are independent of the actual magnification of the camera system.

\subsection*{Analysis of absorption imaging data}

The raw data of the absorption imaging of both detection systems was analyzed in the following way.
For the initial calibration of the atom source generation with short times-of-flight a bimodal fitting procedure was used to consider both the BEC and the thermal clouds. For this purpose a two-dimensional (2D) Thomas-Fermi and a 2D-Gaussian density profile were superimposed and fitted to the absorption data together. In this way the atom numbers, positions, and sizes of the clouds were extracted and information on the BEC fraction was obtained. 

For longer free evolution times, and in particular for the interferometry measurements, the thermal cloud was hard to distinguish from the background noise. Thus, only a 2D-Thomas-Fermi profile was used in these cases to obtain atom numbers, central positions, and Thomas-Fermi radii. 

Furthermore, in case of the differential interferometers used to measure the magnetic field curvature the whole data set was post-processed by a principal component analysis~\cite{Li2007} in order to remove background imaging noise and obtain a better signal-to-noise ratio.

\subsection*{Classical potential curvature measurement}
By performing only a single beam splitting pulse at $t = t_0$ one can observe the trajectories of atoms in the $\zerohk$ and $\twohk$ state simultaneously for long times. 
The differential motion of these two clouds reveals information about the differential velocity $\Delta v = v_{\twohk} - v_{\zerohk}$ as well as the potential curvature $\Gamma$ acting differentially on both clouds, i.e., the \textit{trapping} frequency $\omega=\sqrt{\Gamma}$ of any residual harmonic potential. 
Starting from the classical equations of motion in a harmonic potential and by taking into account that both clouds are at the same position at $t_0$ we obtain the distance
\begin{align}
    \Delta x_j = \frac{\Delta v_j}{\omega_j} \sin\left(\omega_j (t-t_0)\right) \qquad \text{for} \quad t>t_0 
    \label{eq:method:distance_diff_motion}
\end{align}
between both clouds in any spatial dimension $x_j = x,y,z$ and associated velocities $v_j$ and frequencies $\omega_j$.

\begin{figure}[h]
    \centering
    \includegraphics{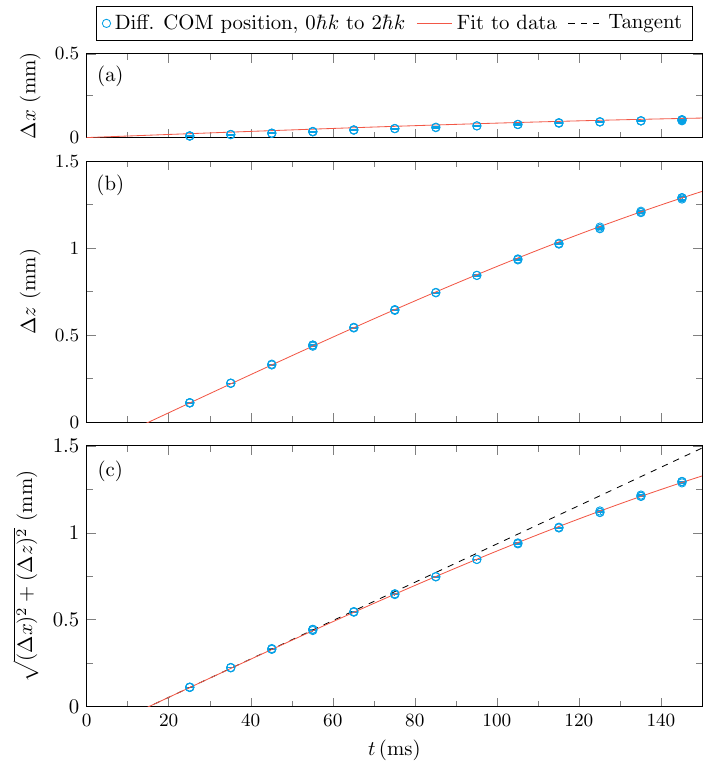}
    \caption{\textbf{ Measurement of the residual trap frequency along the Bragg beam by differential center-of-mass (COM) motion:} (a-c) Spatial separation of the  $\zerohk$ and $\twohk$ momentum states of $^{87}$Rb BECs simultaneously observed during free evolution up to expansion times of 150 ms. After release from the trap a splitting pulse of the Bragg beam is applied at $t= \qty{15}{\milli\second}$ for a duration of $\qty{0.13}{\milli\second}$ generating an equal superposition of both momentum states. 
    (a-b) Comparing the distance over time between both states in the $x$- and $z$-direction allows determination of the angle $\alpha = \qty{4.77 \pm 0.04}{\degree}$ between the Bragg beam and the $z$-axis.
    (c) For the total distance the experimental data (blue dots) clearly deviates from a simple linear behavior (black dashed line) due to spatially dependent forces acting on the atoms. Fitting a sine curve (red line) to the data allows to quantify the residual trap frequency along the direction of the Bragg beam to $\omega_{\mathrm{COM},z^\prime} = 2\pi (0.965 \pm 0.014)$ Hz (see Methods). }
    \label{fig:supplementary_figure_A2}
\end{figure}

The differential motion can be used to measure the residual trapping frequencies $\omega_j$. We have plotted the total distance between the two split $\mftwo$ atom clouds in Fig.~\ref{fig:supplementary_figure_A2} as well as per spatial direction. Using Eq.~\eqref{eq:method:distance_diff_motion}, we obtain $\omega_{\mathrm{COM},z^\prime} = 2\pi \times (0.965 \pm 0.014)$ Hz for the harmonic trapping frequency along the direction of the Bragg beam, or a potential curvature $\Gamma = 36.76 \pm 1.07 \;\si{\per \square \second}$.

Complementary measurements with atoms in the $\mfzero$ state reveal no additional forces acting on the atoms, such that the potential curvature measured for the $\mftwo$ atoms can be attributed to a magnetic field curvature in line with the differential interferometric measurements. 
Thus, the result for the corresponding magnetic field curvature based on the linear Zeeman effect is $\left|B''\right| = \qty{572 \pm 17}{\nano\tesla \per \square \milli \metre}$.

To ensure consistency of the measurement campaign and potential curvature stability over time, we have repeated the differential center-of-mass campaign shown in Fig.~\ref{fig:supplementary_figure_A2} after approximately one year. The results of both campaigns are in excellent agreement with each other.

\subsection*{Ellipse fitting}

For the differential interferometers reported here we have measured the relative populations $N_{\mathrm{rel},j} = N_{0\hbar k, j}/(N_{0\hbar k, j} + N_{2\hbar k, j})$ of the $\zerohk$ states in both MZIs labeled by $j = \text{I}, \text{II}$ through absorption imaging for different laser phase values.
Since the Bragg pulses act simultaneously on both MZIs the corresponding populations can be described by the relations
\begin{align}
\mathcal{N}_{\text{rel,I}}  &= \mathcal{N}_{\text{I}, 0} +  \frac{\mathcal{V}_{\text{I}}}{2} \cos \left( \phi_{\text{I}} \right) \,, \label{eq:method:pop_model_MZI_1} \\
\mathcal{N}_{\text{rel,II}}  &=  \mathcal{N}_{\text{II}, 0} +  \frac{\mathcal{V}_{\text{II}}}{2} \cos \left( \phi_{\text{I}} +  \pi + \Delta \phi \right) \,, \label{eq:method:pop_model_MZI_2}
\end{align}
where $\mathcal{N}_{j,0}$ is the offset population, $\mathcal{V}_{j}$ the visibility and $\Delta \phi = \phi_{\text{II}} - \phi_{\text{I}}$ the differential phase between both MZIs. The additional phase of $\pi$ for the second interferometer is due to the fact that the input for this interferometer is an atom cloud in the momentum state $\twohk$ rather than $\zerohk$.

In a noisy environment such as the ISS, it is hard to extract the phase of a single interferometer independently, which is why we consider the correlated data sets and fit both model functions, Eqs.~\eqref{eq:method:pop_model_MZI_1} and \eqref{eq:method:pop_model_MZI_2}, simultaneously to the data. 
Consequently, there are five fit parameters which for convenience we stack into the parameter vector $\vec{\theta} = \left( \mathcal{N}_{\text{I}, 0}, \, \mathcal{V}_{\text{I} }, \, \mathcal{N}_{\text{II}, 0}, \,\mathcal{V}_{\text{II} }, \, \Delta \phi \right)^T$.

In order to obtain the optimal values for these parameters we perform a nonlinear least square optimization based on the loss function
\begin{align}
L  (\vec{\theta} )  = \frac{1}{N_{\text{data}}} \sum_{k=1}^{N_{\text{data}}} \, \underset{{l=1, ..., N_{\text{disc}}}}{\operatorname{min} } \, \Delta_{k,l}^2  +  \frac{1}{N_{\text{disc}}}\sum_{l=1}^{N_{\text{disc}}} \, \underset{{k=1, ..., N_{\text{data}}}}{\operatorname{min} } \, \Delta_{k,l}^2 \,. \label{eq:method:loss_function}
\end{align}
Here $N_{\text{data}}$ is the number of experimental data points and $N_{\text{disc}}$ is the number of points used to discretize the model functions $\mathcal{N}_{\text{rel,I}}$ and $\mathcal{N}_{\text{rel,II}}$ for equally distributed values $\phi_\text{I} \in [0, 2\pi ]$. 
The distance between the individual experimental data points and the discretized points of the ellipse is given by the matrix entries $\Delta_{k,l}$. 
Hence, the loss function, Eq.~\eqref{eq:method:loss_function}, is determined both by the minimum distance between each experimental data point and the nearest point of the model functions (first sum) as well as the minimum distance between each discretization point and its nearest experimental data point (second sum).

This loss function approach ensures that closely clustered data points are not mistakenly identified as parts of a larger ellipse, particularly when these points cover only a small segment of the fitted ellipse's total circumference. Furthermore, the method takes into account the uneven spacing of measurement points due to the position-dependent curvature of an ellipse. This consideration prevents regions of lower curvature, where points are less densely packed, from being under weighted, thus avoiding an inaccurately narrow fit for the ellipse. This aspect is especially critical for data that is strongly correlated ($\Delta\phi \approx 0$) or anti-correlated ($\Delta\phi \approx \pi$), where there is a risk of the data being erroneously fitted to a very large ellipse representing an effectively straight line in the correlation plot.

Another effect has to be taken into account for strongly correlated ($\Delta\phi \approx 0$) or anti-correlated ($\Delta\phi \approx \pi$) data as well. 
In the ideal case, a vanishing differential phase would lead to data populating a diagonal in the correlation plot. 
However, in the presence of detection noise, the experimental data is spread around this diagonal, such that any ellipse fit would yield a non-zero $\Delta\phi$ value because a slightly opened ellipse fits the data better and minimizes the loss function. 
Hence, it is inherently hard for ellipse fitting techniques to distinguish between truly non-zero differential phase values and vanishing differential phases~\cite{Stockton_2007}.
As a consequence, we extend the error bars to include 0 if the differentiation cannot be done reliably. 
The same holds for values close to $\pi$. 

\begin{figure}[h]
    \centering
    \includegraphics[width=1\textwidth]{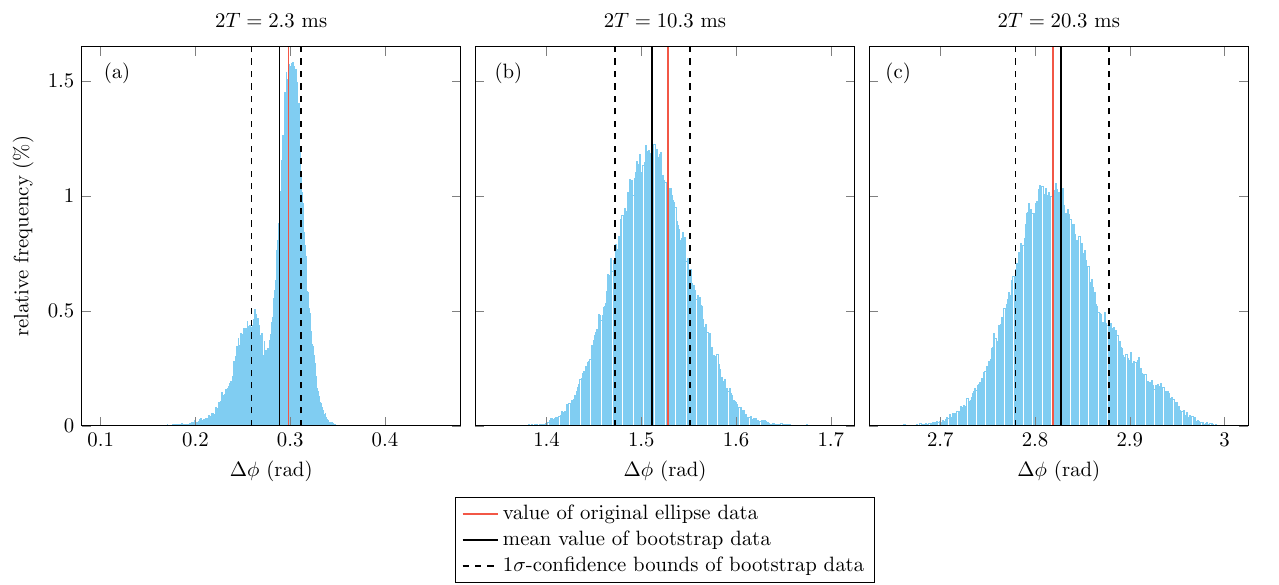}
    \caption{\textbf{Bootstrap analysis of the differential phase for differential Mach-Zehnder atom interferometers:} (a-c) Histograms of the differential phase $\Delta\phi$ obtained by fitting a total of $10^5$ bootstrapped data sets sampled from the corresponding experimental data sets shown in Fig.~\ref{fig:main_figure_3}~(a-c) for interferometer times $2T = 2.3$ ms, $10.3$ ms, and $20.3$ ms (see Methods). 
    The value of $\Delta\phi$ obtained from the original experimental data set (red line) is in all three cases close to the mean value of the bootstrap histogram (black line) indicating the high consistency of the data sets. 
    The $1\sigma$-confidence bounds (black dashed line) are considered as the uncertainty of the phase estimation.
    The histogram in subplot (a) features a bimodal distribution which is caused by the few outliers with very high or low measured population displayed in Fig.~\ref{fig:main_figure_3}~(a), while the histograms in subplot (b) and (c) approximately follow the typical Gaussian behavior.
    }
    \label{fig:supplementary_figure_A3}
\end{figure}

Due to the non-differentiability of the loss function $L(\vec{\theta})$, the Hessian or curvature matrix, which at the optimum parameters $\vec{\theta}^{\ast}$ is proportional to the inverse of the covariance matrix, cannot be obtained analytically. Numerical approximations of the Hessian matrix at the optimum show limited robustness to slight variations in the values of the optimal parameters making the precise determination of error and confidence bounds challenging.
In order to nevertheless estimate the parameter errors properly we perform a bootstrap analysis of the correlated experimental data ${N}_{\text{rel,I}}$ and ${N}_{\text{rel,II}}$. 
For this purpose, we sample a number of $N_{\text{BS}}$ datasets by drawing $N_\text{data}$ data points with replacement from the experimental data. 
Next, we fit the model functions Eqs.~\eqref{eq:method:pop_model_MZI_1} and \eqref{eq:method:pop_model_MZI_2} again to each of the $N_{\text{BS}}$ synthetically generated bootstrap data sets and obtain the optimal parameters in each case. 
The resulting statistical distribution of the five fit parameters is then analyzed for measures like the standard deviation or confidence intervals, where the $1\sigma$-confidence bounds obtained in this way are used as the experimental uncertainties in determining the differential phase $\Delta\phi$.
Exemplary histograms for the bootstrap analysis of differential interferometers with interferometer times $2T = 2.3$ ms, $10.3$ ms, and  $20.3$ ms are shown in Fig.~\ref{fig:supplementary_figure_A3}.

\subsection*{Phase determination including potential contributions beyond second order}

The treatment presented in the main text assumes instantaneous laser pulses and quadratic magnetic potentials.
We have also performed a more general modeling, which we summarize in this section.

On the one hand, we have modeled the effects of finite pulse duration (taking into account the actual Rabi frequencies and two-photon frequency detunings employed for each pulse) by solving the Schr\"odinger equation that governs the evolution of the atomic wave packets during each pulse and includes the optical potential associated with the laser field. Our treatment extends the results~\cite{Antoine_2006} for finite pulse duration.
Furthermore, in order to include the effects of finite pulse durations in treatments for gravity gradients that originally assumed instantaneous pulses, we consider the freely falling frames co-moving with the mid-point trajectory of each interferometer~\cite{Roura_2025}.

On the other hand, we have considered the effects of potential contributions beyond second order.
Here, we show the more complete phase terms and later justify our simplifications.
Including gradients up to third order, we write the potential along the Bragg beam direction $z'$ as
\begin{align}
    V=V_0 + m\left(-a_0\,z'+\frac{\Gamma_0}{2}\,z'^2-\frac{\gamma_0}{6}\,z'^3\right)
    \label{eq:potential_third_order}
\end{align}
with a potential offset $V_0$, the mass $m$ of a $^{87}$Rb atom and the uniform acceleration $a_0$, acceleration gradient $\Gamma_0$ and acceleration curvature $\gamma_0$.
The phase of a Mach-Zehnder interferometer in a quadratic potential has been calculated in a number of references \cite{Antoine_2003,Bongs_2006,Hogan_2008,Roura_2014,Roura_2017} and the result is given by
\begin{align}
\phi _\text{MZI} / k_\text{eff} =
aT^2
-  \left( v_0 + v_\text{rec} \right) \Gamma T^3
- \frac{7}{12} a \Gamma T^4
+ \mathcal{O}\left(\Gamma^2\right)
\label{eq:phase_MZI_second_order}
\end{align}
with the pulse separation time $T$, the single-photon recoil velocity $v_\text{rec}$, and the position (velocity) $z_0$ ($v_0$) of the atoms entering the interferometer.
We now expand up to second order the full potential including the acceleration curvature $\gamma$ for the two spatially separated MZI-I and MZI-II with initial positions $z_0^I, z_0^{II}$ and initial velocities $v_0^I,v_0^{II}$, respectively.
In this description, we choose the initial velocities to describe the lower path for both interferometers and post-process the experimental data to handle the $\pi$ phase shift due to different initial momentum states.
This yields the following local accelerations and acceleration gradients, respectively:
\begin{align}
    a_j = a_0 - \Gamma_0 z_j + \frac{1}{2}\gamma_0 z_j^2, \qquad \Gamma_j  = \Gamma_0 - \gamma_0 z_j.
    \label{eq:expanded_acc_grad}
\end{align}
Inserting those into Eq.~\eqref{eq:phase_MZI_second_order} and calculating the differential phase between MZI-I and -II yields:
\begin{align}
    \Delta \phi _\text{MZI} / k_\text{eff}
    ={}& \phi_\text{MZI-II} - \phi_\text{MZI-I} \label{eq:diff_phase_MZI_expanded} \\
    = {}& \left[- \Delta z \Gamma_0 + \frac{1}{2} \Delta z \, \left( \Delta z + 2 z_0^I\right)\gamma_0 \,\right] T^2 \nonumber\\
&- \left[ \Delta v \, \Gamma_0 -  \left( \Delta z \,(v_0^I + v_\text{rec}) - \Delta v \,(\Delta z + z_0^I) \right) \gamma_0  \,\right] T^3 \nonumber\\
& + \frac{7}{12} \Delta z \, a_0 \gamma_0 T^4 \nonumber\\
&+ \mathcal{O} \left(\Gamma^2 , \gamma^2,\Gamma \gamma \right) \nonumber
\end{align}
with the differential position (velocity) $\Delta z$ ($\Delta v$) between both MZIs.

The butterfly interferometer phase in a quadratic potential is given by~\cite{Kleinert2015}
\begin{align}
\phi _\text{BFI} / k_\text{eff} = 
2 \left( v_0 + v_\text{rec} \right) \Gamma T^3
+ 4 a \Gamma T^4
+ \mathcal{O}\left(\Gamma^2 \right).
\label{eq:phase_BFI_second_order}
\end{align}
Inserting the locally evaluated terms from Eq.~\eqref{eq:expanded_acc_grad} into Eq.~\eqref{eq:phase_BFI_second_order} yields the differential BFI phase:
\begin{align}
    \Delta \phi _\text{BFI} / k_\text{eff}
    ={}& \phi_\text{BFI-II} - \phi_\text{BFI-I} \label{eq:diff_phase_BFI_expanded} \\
    = {}& \frac{1}{4} \left[ \Delta v \, \Gamma_0 -  \left( \Delta z \,(v_0^I + v_\text{rec}) + \Delta v \,(\Delta z + z_0^I) \right) \gamma_0  \,\right] T^3 \nonumber\\
    &- \frac{1}{4} \Delta z \, a_0 \gamma_0 T^4 \nonumber\\
    &+ \mathcal{O} \left(\Gamma^2 , \gamma^2,\Gamma \gamma \right) \nonumber
\end{align}
where, in contrast to some literature, we used the nomenclature from the main part of our paper where one butterfly interferometer takes a total time of $2T$.

The differential butterfly with the longest interferometer time $2T=\qty{20.5}{\milli\second}$ measured a differential phase of $\Delta \phi^\text{BFI} = 0.22 ^{+0.1}_{-0.22} \, \mathrm{rad}$.
As this differential phase result includes zero, we evaluate an upper bound for the acceleration curvature $\gamma_0$.
For this purpose, we require knowledge about the other terms contributing to the differential BFI phase in Eq.~\eqref{eq:diff_phase_BFI_expanded}.
The kinetic properties can be calculated based on our preparation sequence and considering the dynamics determined by Eq.~\eqref{eq:potential_third_order}.
We have quantified the acceleration acting on the $\mftwo$ atoms with a COM time-of-flight measurement as $a_0= (0.11 \pm 0.02)\,\si{\meter \per \square \second}$.
As an approximation for the acceleration gradient, we use the measurement result from the differential MZI with $\mftwo$ atoms, $|\Gamma| = \qty{39.46 \pm 0.02}{\per \square\second}$.
Taking into account all possible signs of $\Gamma_0,\gamma_0$, we obtain a maximal region of compatible third-order potential contributions of $|\gamma_0^\text{\,max}| \le \qty{1.73e4}{\per \meter \square \second}$.

We fitted the complete phase model shown in Eq.~\eqref{eq:diff_phase_MZI_expanded} to our differential MZI data shown in Fig.~\ref{fig:main_figure_4}, using as unconstrained parameters both the acceleration gradient $\Gamma_0$ and the acceleration curvature $\gamma_0$.
Although this complete model is fitting the data well, too, we obtain an inconsistent third-order contribution an order of magnitude larger than our bound $\gamma_0^\text{\,max}$.
We thus conclude that the experimental data spread shown in Fig.~\ref{fig:main_figure_5} when using a purely second-order model cannot be sensibly explained by higher-order contributions.
To avoid incorrectly fitting unknown effects or drifts in the potential landscape, we refrain from using third-order potential contributions.
We make a final simplification to neglect the term $\Delta v \Gamma_0 T^3 \ll \Delta z \Gamma_0 T^2$ being more than two orders of magnitude smaller for our experimental parameters.

\clearpage

\section*{Data Availability Statement}
The datasets generated for and analyzed in this paper are available from the corresponding author upon reasonable request. 
All NASA CAL data are on a schedule for public availability through the NASA Physical Science Informatics (PSI) website (\url{https://www.nasa.gov/PSI}).

\section*{References}

\bibliography{BibFile}

\begin{acknowledgments}
We gratefully acknowledge the contributions of current and former members of CAL’s operations and technical teams, and those of the team at Infleqtion.
M.M. acknowledges helpful discussions with Cass Sackett about ellipse fitting of differential interferometer data.
G.M. acknowledges helpful discussions with Christian Struckmann and Michael Werner about interferometer phases beyond quadratic potentials.
This work is supported by the Biological and Physical Sciences division of NASA’s Science Mission Directorate at the agency’s headquarters in Washington D.C. and by the ISS Program Office at NASA’s Johnson Space Center in Houston TX, through RSA No. 1616833, and the DLR Space Administration with funds provided by the Federal Ministry for Economic Affairs and Climate Action (BMWK) under grant numbers 50WM2245-A/B (CAL-II), 50WM2545-A/B (CAL-III) and 50WM2263A (CARIOQA-GE). N.G., E.M.R. and T.E. gratefully acknowledge financial support from the Deutsche Forschungsgemeinschaft (DFG) through SFB 1227 (DQ-mat) within Project A05, Germany’s Excellence Strategy (EXC-2123 QuantumFrontiers Grants No. 390837967). Cold Atom Lab was designed, managed, and operated by the Jet Propulsion Laboratory, California Institute of Technology, under contract with the National Aeronautics and Space Administration (Task Order 80NM0018F0581). A.R.\ is supported by the Q-GRAV and AICEF Projects within the Space Research and Technology Program of the German Aerospace Center (DLR).
\end{acknowledgments}

\section*{Author contributions}
\textsuperscript{\dag}These authors contributed equally: Matthias Meister, Gabriel Müller

M.M., G.M., P.B., A.P., and J.S. analyzed the atom source and classical motion, and performed the Bragg beam characterization. D.B.R. together with M.M. performed the ellipse fitting and bootstrap analysis. M.M., G.M., and A.R., together with P.B., T.E., and E.G., and supported from H.A., W.H., C.S., and E.C., analyzed the differential interferometer data. 
A.R. performed a detailed numerical modelling of the interferometers.
J.R.W. communicated the consortium’s sequences to the ISS and executed the science campaigns. M.M., H.M., E.M.R., W.P.S., and N.G. are co-investigators and N.P.B. is the principal investigator and director of the CUAS consortium. The initial manuscript was drafted by M.M., and G.M. together with D.B.R., T.E., J.S., N.G., and N.P.B.. All authors read, edited and approved the final manuscript.

\end{document}